\documentclass{article}
\usepackage{spconf,amsmath,graphicx,bm,amssymb}
\usepackage{algorithm}
\usepackage{algorithmic}

\renewcommand{\frac}{\dfrac}

\newtheorem{dingli}{Theorem~}

\title{An Efficient Active Set Algorithm for Covariance Based Joint Data and Activity Detection for Massive Random Access with Massive MIMO}
\name{Ziyue Wang$^{\star,\S}$, Zhilin Chen$^{\dag}$, Ya-Feng Liu$^{\S}$, Foad Sohrabi$^{\dag}$, and Wei Yu$^{\dag}$}

\address{$^{\star}$School of Mathematical Sciences, University of Chinese Academy of Sciences, Beijing, China\\[2pt]
    $^{\dag}$ Department of Electrical and Computer Engineering, University of Toronto, Toronto, Canada\\[2pt]
    $^{\S}$LSEC, ICMSEC, AMSS, Chinese Academy of Sciences, Beijing, China \\[2pt]
  Email: wangziyue20@mails.ucas.ac.cn,  \{zchen, fsohrabi, weiyu\}@ece.utoronto.ca, yafliu@lsec.cc.ac.cn}
\begin{document}
\ninept
\maketitle
%
\begin{abstract}
This paper proposes a computationally efficient algorithm to solve the joint data and activity detection problem for massive random access with massive multiple-input multiple-output (MIMO). The BS acquires the active devices and their data by detecting the transmitted preassigned nonorthogonal signature sequences. This paper employs a covariance based approach that formulates the detection problem as a maximum likelihood estimation (MLE) problem. To efficiently solve the problem, this paper designs a novel iterative algorithm with low complexity in the regime where the device activity pattern is sparse -- a key feature that existing algorithmic designs have not previously exploited for reducing complexity. Specifically, at each iteration, the proposed algorithm focuses on only a small subset of all potential sequences, namely the \emph{active set}, which contains a few most likely active sequences (i.e., transmitted sequences by all active devices), and performs the detection for the sequences in the active set. The active set is carefully selected at each iteration based on the current detection result and the first-order optimality condition of the MLE problem.
Simulation results show that the proposed active set algorithm enjoys significantly better computational efficiency (in terms of the CPU time) than the state-of-the-art algorithms.




%
%
\end{abstract}
\begin{keywords}
Active set, joint data and active detection, massive MIMO, massive random access.
\end{keywords}
\section{Introduction}
\label{sec:intro}
Massive machine-type communication (mMTC) is a main use case in the fifth generation
(5G) cellular systems \cite{Bockelmann2016}. A challenging task in mMTC is the uncoordinated random access, in which a large number of sporadically active devices wish to send small data to the base-station (BS) in the uplink \cite{ChenX2020}. To meet the low-latency requirement in mMTC, the grant-free random access scheme could be a promising solution \cite{Liu2018b,Senel2018}, in which each device is preassigned multiple signature sequences from a large set of nonorthogonal sequences, and the active device selects one sequence from the assigned sequences to transmit. The data and device identification are embedded in the sequence selection. The BS then detects the active devices and decodes their data by detecting which sequences are transmitted.


By exploiting the sporadic device traffic, the joint data and activity detection problem has been formulated as a compressed sensing problem \cite{Senel2018}, in which the data and the device activity are recovered along with the instantaneous channel state information (CSI) via the approximate message passing (AMP) algorithm. Similar methods have also been used for the scenario where each device is associated with only one sequence for the purpose of device activity detection \cite{Liu2018,Chen2018,liu2021efficient}. However, if the CSI is not needed, it is actually possible to recover the data and activities without recovering the channel coefficients using a convariance based approach
 \cite{zhilin_icc2019,ChenZ2020}, which outperforms the AMP method, especially in the massive multiple-input multiple-output (MIMO) systems. In the covariance based method, the detection problem is formulated as a maximum likelihood estimation (MLE) problem, in which the channel coefficients are treated as random samples and averaged out in computing the covariance. This covariance based method is first suggested in \cite{Haghighatshoar2018} for device activity detection, and it has also been used for a few related data/activity detection problems, e.g., data decoding for unsourced random access \cite{Fengler2019}, cooperative activity detection in cell-free systems \cite{Shao2020a}, and activity detection with interference \cite{Jiang2020a}.

The coordinate descent (CD) algorithm that iteratively updates the sequence selection for each device is commonly used in solving the detection problem in the covariance based approach, which achieves excellent detection performance; see \cite{zhilin_icc2019,Haghighatshoar2018,Jiang2020a} for more details. The possible reason for the popularity of the CD algorithm is that its subproblem (i.e., the original problem with respect to only one variable) admits a nice closed-form solution \cite{Haghighatshoar2018}, which makes it easily implementable. To further speed up the convergence of the CD algorithm, a new coordinate sampling strategy is proposed in \cite{Dong2020}. Other algorithms for solving the detection problem include the expectation maximization/minimization (EM) algorithm (i.e., sparse Bayesian learning) \cite{Wipf2007} and the SPICE algorithm \cite{yang2018sparse}. However, none of the above mentioned solutions take advantage of the sparsity of the true solution of the detection to lower their algorithmic complexities, thus becoming less computationally efficient when the problem's dimension is huge, which is the case in mMTC.

In this paper, we propose a computationally efficient algorithm that carefully exploits the sparsity of the true solution to solve the joint data and activity detection problem in the covariance based approach. Specifically, we propose an iterative algorithm that attacks the original large-scale problem by solving a sequence of small-size problems. We focus on only a small subset of all sequences at each iteration, termed as the \emph{active set}, which contains only the most likely active sequences\footnote{Active sequences in this paper refer to transmitted sequences by all active devices in the joint data and activity detection problem.} and can be seen as an approximation of the set of active sequences. We perform joint data and activity detection for only the sequences in the active set using a low-complexity spectral projected gradient (PG) algorithm \cite{birgin2000nonmonotone}. We carefully update the active set at each iteration based on the current detection result and the first-order optimality condition of the joint detection problem. We also establish the convergence of the proposed active set algorithm. Simulation results show that as compared to the commonly used CD algorithm in the covariance based approach, the proposed active set algorithm has much higher computational efficiency (in terms of the CPU time).

\section{System Model and Problem Formulation}

\subsection{System Model}
Consider an uplink single-cell system where there are one BS equipped with $M\gg 1$ antennas and $N$ devices each equipped with a single antenna. Assume a quasi-static narrow-band channel model, where the wireless channels remain unchanged within each transmission block but may vary over different blocks. Let $\sqrt{g_{n}}\mathbf{h}_{n}\in \mathbb{C}^{M\times 1}$ denote the channel vector from device $n$ to the BS, where $g_n\geq 0$ is the large-scale fading component (depending on the device's location), and $\mathbf{h}_{n}\in \mathbb{C}^{M\times 1}$ is the Rayleigh fading component following the i.i.d.\ complex Gaussian distribution. 

In each coherence block, only $K\ll N$ devices are active (due to the sporadic traffic), and each active device wishes to transmit $J$ bits of data to the BS, where $J$ is a small value in the mMTC scenario. Assume that each device $n$ has a unique signature sequence set $\mathcal{S}_n=\left\{\mathbf{s}_{n,1},\mathbf{s}_{n,2}, \ldots, \mathbf{s}_{n,Q}\right\},$ where~
 $\mathbf{s}_{n,q} \in \mathbb{C}^{L\times 1}, 1\leq q\leq Q\triangleq 2^J,$ and $L$ is the signature sequence length. When device $n$ is active and needs to send $J$ bits {of} data, it selects one sequence from $\mathcal{S}_n$ to transmit. Finally, let $\chi_{n,q}\in\{0,1\}$ indicate whether or not sequence $q$ of device $n$ (i.e., $\mathbf{s}_{n,q}$) is transmitted. Notice that at most one sequence is selected by each device, then it follows that $\chi_{n,q}$ satisfies $\sum_{q=1}^Q\chi_{n,q}\in\{0,1\}$, where $\sum_{q=1}^Q\chi_{n,q}=0$ indicates that device $n$ is inactive, and $\sum_{q=1}^Q\chi_{n,q}=1$ indicates that device $n$ is active.

Assume that the {sequences transmitted by active} devices are perfectly synchronized. Then the received signal $\mathbf{Y}\in \mathbb{C}^{L\times M}$ at the BS, which is a superposition of the
transmitted signals from all active devices, can be expressed as
\begin{align}\label{eq.sys}
\mathbf{Y}&=\sum_{n=1}^{N}\sum_{q=1}^{Q}\chi_{n,q}\mathbf{s}_{n,q}\sqrt{g_n}\mathbf{h}_{n}^T +\mathbf{W},
\end{align}
where $\mathbf{W}\in \mathbb{C}^{L\times M}$ is the effective
i.i.d.\ Gaussian noise whose variance $\sigma_{w}^2$ is the background noise power normalized by the device transmit power.

To obtain a more compact expression of the received signal in \eqref{eq.sys}, we define $\mathbf{S}_n=[\mathbf{s}_{n,1},\ldots,\mathbf{s}_{n,Q}]\in\mathbb{C}^{L\times Q}$,  $\mathbf{D}_n=\sqrt{g_n}\operatorname{diag}\{\chi_{n,1},\ldots,\chi_{n,Q}\}\in\mathbb{C}^{Q \times Q},$ $\mathbf{H}_n=[\mathbf{h}_{n},\ldots,\mathbf{h}_{n}]^T\in\mathbb{C}^{Q \times M}$ for all $n.$ Based on them, we further define $\mathbf{S}=[\mathbf{S}_1,\ldots,\mathbf{S}_N]\in\mathbb{C}^{L\times NQ}$,  $\boldsymbol{\Gamma}^{{1}/{2}}=\operatorname{diag}\{\mathbf{D}_1,\ldots,\mathbf{D}_N\}\in\mathbb{C}^{NQ \times NQ},$
and $\mathbf{H}=[\mathbf{H}_{1}^T,\ldots,\mathbf{H}_{N}^T]^T\in\mathbb{C}^{NQ\times M}$. Then, the received signal in \eqref{eq.sys} can be compactly expressed as
\begin{align}\label{eq.sys.comp}
\displaystyle \mathbf{Y}=\mathbf{S}\boldsymbol{\Gamma}^{{1}/{2}}\mathbf{H}+\mathbf{W}.
\end{align}
Let $\boldsymbol{\gamma}\in\mathbb{C}^{NQ \times 1}$ denote the diagonal entries of $\boldsymbol{\Gamma}$, i.e., $\boldsymbol{\gamma}=[\boldsymbol{\gamma}_1^T,\ldots,\boldsymbol{\gamma}_N^T]^T$, where $\boldsymbol{\gamma}_n=[\gamma_{n,1},\ldots,\gamma_{n,Q}]^T\in\mathbb{C}^{Q \times 1}$ with $\gamma_{n,q}=g_n\chi_{n,q}$. In the following, we will use $\boldsymbol\gamma$ and $\boldsymbol{\Gamma}$ interchangeably.

\subsection{Problem Formulation}
 The joint activity and data detection problem is to detect the variables $\gamma_{n,q}$'s, which indicate both the activity of device $n$ and its data (if it is active) from the received signal $\mathbf{Y}$ based on the knowledge of the signature sequence matrix $\mathbf{S}.$ Specifically, if $\gamma_{n,q}>0,$ then device $n$ is active and it transmits sequence $\mathbf{s}_{n,q};$ otherwise device $n$ is inactive.

 As shown in \cite{Haghighatshoar2018,zhilin_icc2019}, the above joint activity and data detection problem can be mathematically formulated as the MLE problem. Specifically, it can be observed from \eqref{eq.sys.comp} that given $\boldsymbol{\gamma}$, each column of $\mathbf{Y}$, denoted as
 $\mathbf{y}_m\in\mathbb{C}^{L \times 1}, 1\leq m\leq M $,
 can be seen as independent samples from a complex Gaussian distribution as
 \begin{align}\label{eq.gauss}
 \mathbf{y}_m \sim \mathcal{CN}\left(\mathbf{0},\mathbf{S}\boldsymbol{\Gamma}^{1/2}\boldsymbol{\Lambda}\boldsymbol{\Gamma}^{1/2}\mathbf{S}^H+\sigma_w^2\mathbf{I}\right),
 \end{align}
 {where the covariance matrix is obtained by computing $\mathbb{E}[\mathbf{y}_m\mathbf{y}_m^H]$ based on \eqref{eq.sys.comp}}, $\boldsymbol{\Lambda}$ is a block diagonal matrix with each block being the all-one matrix $\mathbf{E}\in \mathbb{R}^{Q\times Q}$, and $\mathbf{I}$ is an identity matrix. Since there is at most one non-zero entry in each diagonal block $\mathbf{D}_n$ in $\boldsymbol{\Gamma}^{1/2},$ the covariance matrix in \eqref{eq.gauss} can be simplified as $$\mathbf{S}\boldsymbol{\Gamma}^{1/2}\boldsymbol{\Lambda}\boldsymbol{\Gamma}^{1/2}\mathbf{S}^H+\sigma_w^2\mathbf{I}
 	=\mathbf{S}\boldsymbol{\Gamma}\mathbf{S}^H+\sigma_w^2\mathbf{I}.$$
Given $\boldsymbol{\gamma},$ we have $p(\mathbf{Y}|\boldsymbol{\gamma})=\Pi_{m=1}^M p(\mathbf{y}_m|\boldsymbol{\gamma}).$ Based on this and \eqref{eq.gauss}, the minimization of $- \tfrac{1}{M} \log p(\mathbf{Y}|\boldsymbol{\gamma})$, equivalent to the maximization of $p(\mathbf{Y}|\boldsymbol{\gamma}),$ can be formulated as
\begin{subequations}\label{eq.prob1}
	\begin{alignat}{2}\label{eq.prob1.1}
	&\underset{\boldsymbol\gamma}{\operatorname{min}}    &\quad&  \log\left|\mathbf{S}\boldsymbol{\Gamma}\mathbf{S}^H+\sigma_w^2\mathbf{I}\right|+\operatorname{Tr}\left(\left(\mathbf{S}\boldsymbol{\Gamma}\mathbf{S}^H+\sigma_w^2\mathbf{I}\right)^{-1}\hat {\boldsymbol\Sigma}\right)\\
	&\operatorname{s.t.} &      & \boldsymbol{\gamma} \geq \mathbf{0},
	\end{alignat}
\end{subequations}
where 
$\hat {\boldsymbol\Sigma}=\mathbf{Y}\mathbf{Y}^H/M$ is the sample covariance matrix computed by averaging over different antennas, and $\boldsymbol{\gamma} \geq \mathbf{0}$ is due to the fact that $\gamma_{n,q}=g_n\chi_{n,q} \geq 0 $ for all $n$ and $q.$ Since the objective function in problem \eqref{eq.prob1} depends on $\mathbf{Y}$ only through the sample covariance matrix $\hat {\boldsymbol\Sigma},$ the approach of estimating activity and associated data based on solving problem \eqref{eq.prob1} is called the covariance based approach. It is worthwhile mentioning that problem \eqref{eq.prob1} reduces to the activity detection problem in \cite{Haghighatshoar2018} if each device has only a single signature sequence (i.e., $J=0$ and thus $Q=1$).

Let $f(\bm{\gamma})$ denote the objective function of problem \eqref{eq.prob1}. Then, for any $q=1,2,\ldots,Q, n=1,2,\ldots,N,$ the gradient of $f(\bm{\gamma})$ with respect to $\gamma_{n,q}$ is
\begin{equation*} \label{gradienteq}
\left[\nabla f(\bm{\gamma})\right]_{n,q} = \textbf{\textup{s}}_{n,q}^H \bm{\Sigma}^{-1} \textbf{\textup{s}}_{n,q} - \textbf{\textup{s}}_{n,q}^H \bm{\Sigma}^{-1} \bm{\hat{\Sigma}} \bm{\Sigma}^{-1} \textbf{\textup{s}}_{n,q}.
\end{equation*}
The first-order (necessary) optimality condition of problem \eqref{eq.prob1} is
\begin{align}\label{eq.th}
\left[\nabla f(\bm{\gamma})\right]_{n,q}
\begin{cases}
=0, & \operatorname{if}~\gamma_{n,q} >0;   \\
\geq 0, & \operatorname{if}~\gamma_{n,q} = 0,
\end{cases}\ \forall~q, n,
\end{align}
which is equivalent to
$$[\bm{\gamma} - \nabla f(\bm{\gamma})]_+ - \bm{\gamma} = \textbf{0},$$ where $[\cdot]_+$ denotes the projection operator onto the nonnegative orthant. It can be checked that computing $\nabla f(\bm{\gamma})$ has a complexity of ${\cal{O}}(NQL^2).$

\section{Proposed Active Set Algorithm}

The basic idea of the proposed active set algorithm for solving problem \eqref{eq.prob1} is to fully exploit the sparsity of its true solution in the algorithmic design, which is in sharp contrast to all existing algorithms such as EM \cite{Wipf2007}, CD \cite{Haghighatshoar2018,zhilin_icc2019}, and SPICE \cite{yang2018sparse}. More specifically, at each iteration, the active set algorithm first judiciously selects an active set then solves the subproblem defined over the variables in the active set with all the other variables fixed being zero. Since the true solution of problem \eqref{eq.prob1} is sparse, it is expected that the cardinality of the carefully selected active set, i.e., the dimension of the subproblem, will be significantly less than the total number of variables of the original problem \eqref{eq.prob1}. Therefore, solving the subproblem defined over the variables in the active set will be much more computationally efficient than directly solving the original problem \eqref{eq.prob1} (over all variables).

\textbf{Selecting the active set.}  In principle, a desirable active set should contain the indices of active sequences in order to correctly detect the active users and associated data; on the other hand, its cardinality should be as small as possible in order to avoid unnecessary computation and improve the computational efficiency.
Our selection strategy of the active set $\mathcal{A}^k$ at a given feasible point $\bm{\gamma}^k$ is based on (i) the engineering insight of the joint activity and data detection problem and (ii) the first-order necessary optimality condition \eqref{eq.th} of the joint detection problem. In particular, for any given feasible point $\bm{\gamma}^k,$ the selected active set $\mathcal{A}^k$ contains the indices whose corresponding values of $\bm{\gamma}^k$ are positive and large (based on (i)) and the indices whose corresponding values of $\nabla f(\bm{\gamma}^k)$ are negative and small (due to (ii)).
Mathematically, the proposed selection strategy of the active set $\mathcal{A}^k$ is
\begin{equation}\label{selectionrule}
	\mathcal{A}^k = \Big\{ (i,q) \mid \gamma^k_{i,q} > \omega_k \ \textup{or} \ [\nabla f(\bm{\gamma}^k)]_{i,q} < -\nu_k \Big\},
\end{equation}
where $\omega_k$ and $\nu_k$ are two positive parameters. {The choices of the parameters $\omega_k$ and $\nu_k$ provide a trade-off between reducing the cardinality of the active set and not missing the active sequences. In general, the smaller these two parameters, the larger the cardinality of the selected active set and the lower probability of missing the active sequences. To make sure of not missing the active sequences, we let $\omega_k\downarrow 0$ and $\nu_k \downarrow 0$ in \eqref{selectionrule}, which means that $\omega_k$ and $\nu_k$ monotonically decrease and converge to zero.}

\textbf{Solving the subproblem.} At the $k$-th iteration, once the active set $\mathcal{A}^k$ is selected, we solve the following subproblem
\begin{subequations}\label{problemAk}
	\begin{alignat}{2}
	\min & \quad \hat f(\bm{\gamma}_{\mathcal{A}^k})\\
	\operatorname{s.t.}& \quad \bm{\gamma}_{\mathcal{A}^k} \geq \mathbf{0},
	\end{alignat} 
\end{subequations} where $\bm{\gamma}_{\mathcal{A}^k}$ is the subvector of $\bm{\gamma}$ indexed by $\mathcal{A}^k$ and $\hat f(\bm{\gamma}_{\mathcal{A}^k})$ is $f(\bm{\gamma})$ defined over $\bm{\gamma}_{\mathcal{A}^k}$ with all the other variables fixed being zero. Obviously, problem \eqref{problemAk} is different from problem \eqref{eq.prob1}. For instance, problem \eqref{problemAk} is defined over $\bm{\gamma}_{\mathcal{A}^k}$, whereas problem \eqref{eq.prob1} is defined over $\bm{\gamma}$. Therefore, the dimension of problem \eqref{problemAk} is potentially much smaller than that of problem \eqref{eq.prob1} (if the set $\mathcal{A}^k$ in \eqref{problemAk} is properly chosen).

We apply the spectral PG algorithm \cite{birgin2000nonmonotone} to solve the subproblem in \eqref{problemAk} until $\bm{\gamma}_{\mathcal{A}^k}^{k+1}$ satisfying
\begin{equation}\label{subproblemtermination}\left\|\left[[\bm{\gamma}_{\mathcal{A}^k}^{k+1} - \nabla \hat f(\bm{\gamma}_{\mathcal{A}^k}^{k+1})]_+ - \bm{\gamma}_{\mathcal{A}^k}^{k+1}\right]\right\| < \varepsilon_k\end{equation} is found, where $\varepsilon_k>0$ is the solution tolerance at the $k$-th iteration. The spectral PG algorithm \cite{birgin2000nonmonotone} is an PG algorithm with the spectral stepsizes, also called the Barzilai-Borwein (BB) stepsizes \cite{bb88}, which approximately solves the Quasi-Newton equation. In the PG algorithm, we need to compute the gradient of the objective function $\hat f(\bm{\gamma}_{\mathcal{A}^k})$ (equivalent to computing the partial gradient of function $f(\bm{\gamma})$ with respect to the variables in $\mathcal{A}^k$), which has a complexity of ${\cal{O}}(\left|{\mathcal{A}^k}\right|L^2).$
Two distinctive advantages of the spectral PG
algorithm \cite{birgin2000nonmonotone} in the context of solving problem \eqref{problemAk} are as follows.
First, the non-negative constraint is easy to project onto, and thus the algorithm can be easily implemented to solve problem \eqref{problemAk}. Second, the algorithm enjoys
a quite good numerical performance due to the use of the alternating BB stepsizes \cite{bb88,dai2005projected}.


Now, we are ready to present the proposed active set PG algorithm for solving problem \eqref{eq.prob1}. The pseudocodes of the proposed algorithm are given in Algorithm \ref{alg:actset}.


\begin{algorithm}[H]
	\caption{Proposed active set PG algorithm for solving problem \eqref{eq.prob1}}
	\begin{algorithmic}[1]
		\STATE \textbf{Initialize:} $\bm{\gamma}^0 = \mathbf{0}$, $k = 0$, $\left\{\omega_k, \nu_k, \varepsilon_k\right\}_{k\geq0},$ and $\varepsilon > 0;$
		\REPEAT
		\STATE Select the active set $\mathcal{A}^k$ according to \eqref{selectionrule}; 
		\STATE Apply the spectral PG algorithm \cite{birgin2000nonmonotone} to solve the subproblem \eqref{problemAk} until \eqref{subproblemtermination} is satisfied;
		\STATE 
		Set $k \leftarrow k+1;$
		\UNTIL{$\|[\bm{\gamma}^k - \nabla f(\bm{\gamma}^k)]_+ - \bm{\gamma}^k\| < \varepsilon$}
		\STATE \textbf{Output:}  $\bm{\gamma}^k$
	\end{algorithmic} \label{alg:actset}
\end{algorithm}

Next, we present some convergence properties of the proposed active set PG Algorithm \ref{alg:actset} (without rigorous proofs due to the space limitation). Note that a not careful selection of the active set (and parameters in it) might lead to oscillation (and divergence) of the corresponding active set algorithm among different active sets. The following finite termination property is mainly because of the activity set selection strategy in \eqref{selectionrule} (and careful choices of parameters $\omega_k$ and $\nu_k$) and the convergence property of the spectral PG algorithm.

\begin{dingli}\label{theorem}
	For any given tolerance $\varepsilon>0,$ suppose that the parameters $\omega_k$ and $\nu_k$ in \eqref{selectionrule} satisfy $\omega_k\downarrow 0$ and $\nu_k \downarrow 0$ and the parameter $\varepsilon_k$ in \eqref{subproblemtermination} satisfy $\lim\limits_{k \rightarrow \infty}\varepsilon_k < \varepsilon,$ then the active set PG Algorithm~1 will terminate within a finite number of iterations.
\end{dingli}



\section{Simulation Results}

In this section, we present some simulation results to show the efficiency of the proposed active set PG algorithm for solving the joint data and activity detection problem \eqref{eq.prob1}. We generate the same parameters as in \cite{zhilin_icc2019} in our numerical simulations. More specifically, we consider a single cell of radius 1000m and consider the worst-case scenario where all devices are located in the
cell edge such that the large-scale fading components $g_n$'s are the same for all devices. The power spectrum density of the background noise is -169dBm/Hz over 10 MHz and the transmit power of each device is set as 25dBm. The number of BS antennas, the length of the signature sequence, and the bits of the data are set to be $M=256,$ $L=150,$ and $J=1$ (and thus $Q=2$), respectively. We generate all signature sequences from i.i.d.\ complex Gaussian distribution
with zero mean and unit variance.
We set $K/N=0.1,$ which means that $10\%$ of the total devices are active, and compare the performance of different algorithms as $N$ increases.
The parameters in the proposed Algorithm \ref{alg:actset} are
\begin{align}
\omega_k&={10^{-6-k} },\varepsilon_k=\max\left\{10^{-k},0.8*10^{-3}\right\},\nonumber\\
\nu_k& =
\min\left\{10^{4-k},0.5\left|\min_{n,q}\left\{\left[\nabla f(\bm{\gamma}^k)\right]_{n,q}\right\}\right|\right\},\nonumber
\end{align}
and $\varepsilon=10^{-3}.$ All simulation results in this section are obtained by averaging over $500$ Monte-Carlo runs.

\begin{figure}[t]
	\includegraphics[width=0.45\textwidth]{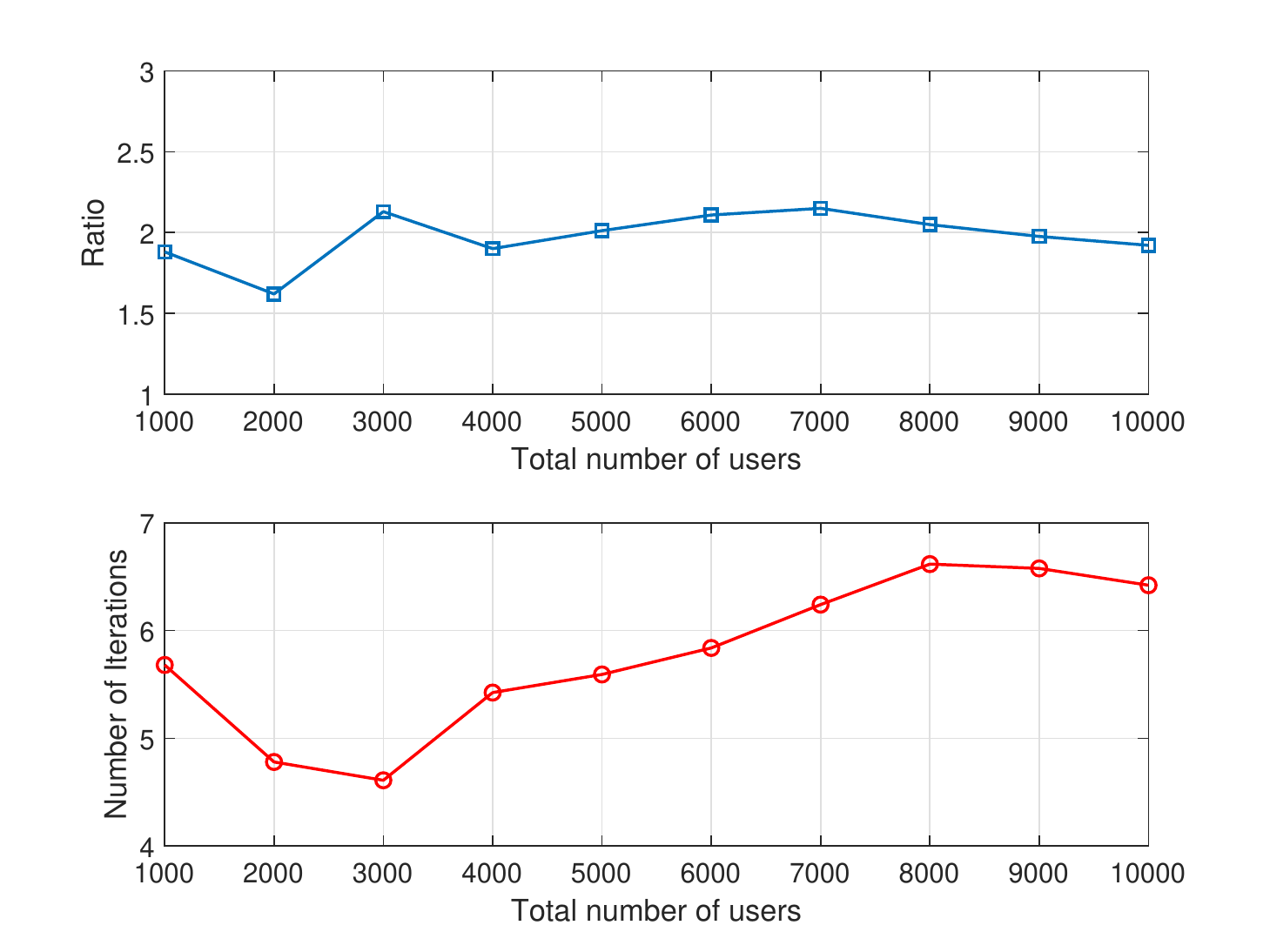}
	\caption{Performance of the proposed active set PG algorithm.}
	\label{ratio}
\end{figure}

The upper subfigure in Fig.~\ref{ratio} plots the average ratio of the cardinality of the selected active sets during all iterations of the proposed Algorithm \ref{alg:actset} and the number of active sequences (i.e., $K$). In the ideal case where the set of active sequences is always selected as the active set in Algorithm 1, the corresponding ratio is $1$; in the worst case where the whole set is always selected as the active set in Algorithm 1, the corresponding ratio is $QN/K=20$. This ratio measures the efficiency of the corresponding active set selection strategy and the smaller the ratio the better the active set selection strategy.
We can observe from the upper subfigure in Fig.~\ref{ratio} that the ratio is in the interval $[1.5, 2.5]$ (and in fact very close to $2$ for different $N$'s), which clearly shows that the proposed active set selection strategy \eqref{selectionrule} is very efficient. The lower subfigure in Fig.~\ref{ratio} plots the average number of iterations that the proposed algorithm needs to terminate. This subfigure shows that Algorithm \ref{alg:actset} will generally terminate within 4--7 iterations, which validates the finite termination result in Theorem \ref{theorem}.


\begin{figure}[t]
	\includegraphics[width=0.45\textwidth]{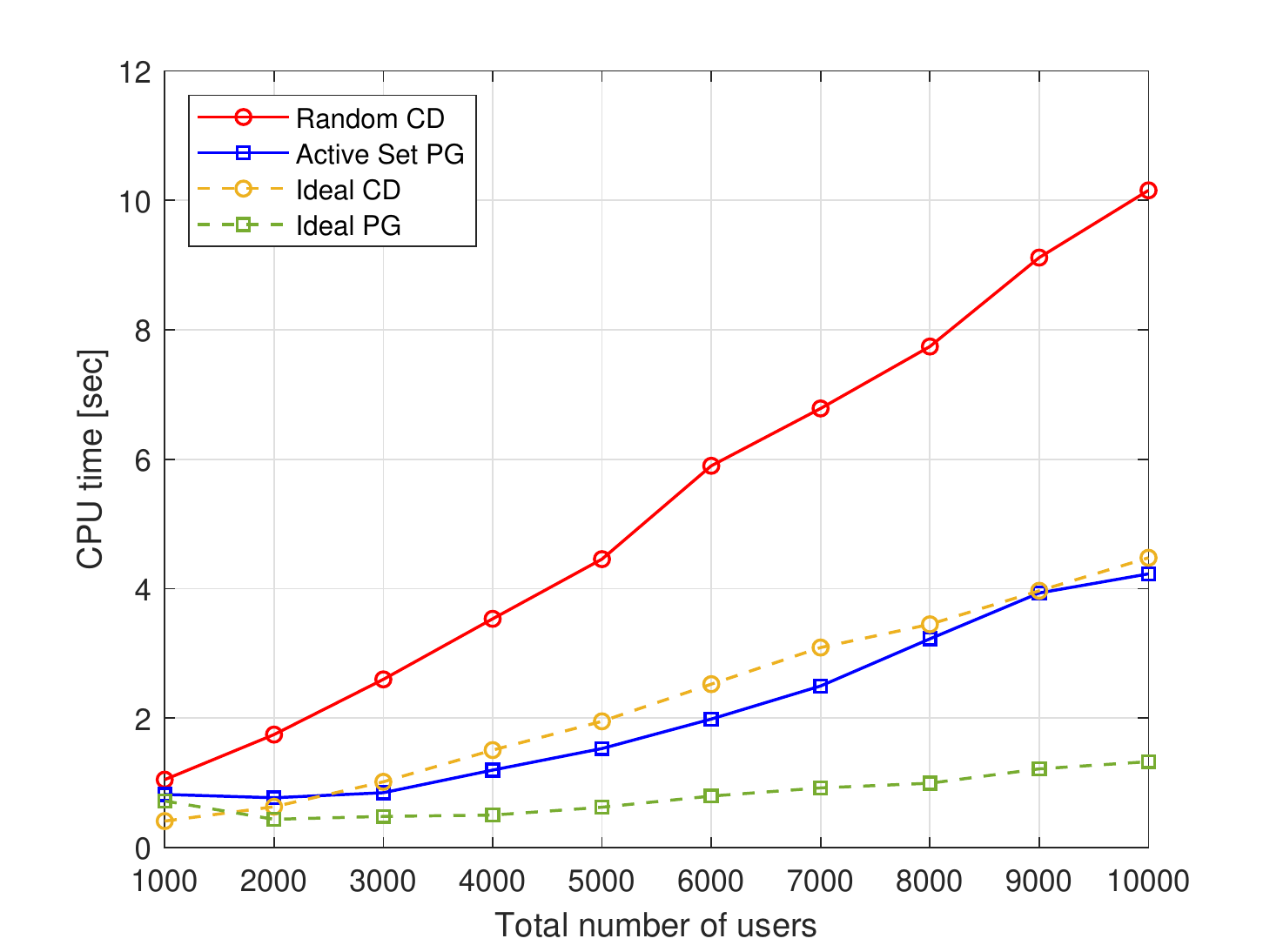}
	\caption{Average CPU time comparison of the proposed active set PG algorithm, the random CD algorithm, the ideal CD algorithm, and the ideal PG algorithm with $K/N=0.1.$}
	\label{fig_time}
\end{figure}

Next, we compare the proposed active set PG Algorithm 1 with the following three benchmark algorithms:
\begin{itemize}
	\item Random CD \cite{zhilin_icc2019,Haghighatshoar2018}: To the best of our knowledge, random CD is the state-of-the-art algorithm for solving the covariance based detection problem \eqref{eq.prob1}, which is much faster than EM \cite{Wipf2007} and SPICE \cite{yang2018sparse}. Among various variants, one of the most efficient ones is the so-called random permuted CD \cite{zhilin_icc2019}, which randomly permutes the indices of all variables at each iteration and then updates the variables one by one according to the order in the permutation in closed form (see line 5 of \cite[Algorithm 1]{Haghighatshoar2018}).
	\item Ideal CD: This refers to the algorithm which applies the random permuted CD algorithm to solve problem \eqref{eq.prob1} defined over the variables in the (ideal) set of active sequences. Since this ideal set is not known at the BS in practice, ideal CD is not a practical algorithm. Ideal CD is compared here for the purpose of characterizing the best possible performance of the CD types of algorithms.
	\item Ideal PG: This refers to the algorithm which applies the PG algorithm \cite{birgin2000nonmonotone} to solve problem \eqref{eq.prob1} defined over the variables in the (ideal) set of active sequences. This algorithm is also not practical and is only theoretically interesting for characterizing the best possible performance of the PG algorithm.
\end{itemize}

We have observed in the simulations that random CD and active set PG algorithms always find the same solution of problem \eqref{eq.prob1}, and thus below we focus on their CPU time comparison. Fig.~\ref{fig_time} plots the CPU time comparison of the proposed algorithm with the above three benchmark algorithms. It can be clearly observed from Fig.~\ref{fig_time} that the proposed active set PG algorithm significantly outperforms the state-of-the-art random CD algorithm \cite{Haghighatshoar2018,zhilin_icc2019} in terms of the CPU time. The proposed algorithm even achieves slightly better computational efficiency than the ideal CD algorithm. 
This shows the importance of exploiting the sparsity of the true solution in order to efficiently solve problem \eqref{eq.prob1}. Note that we fix $K/N=0.1$ in Fig.~\ref{fig_time}. It is expected that the proposed active set PG algorithm will become more efficient than the random CD algorithm as the ratio $K/N$ becomes smaller (i.e., the solution of problem \eqref{eq.prob1} becomes more sparse).


We have also observed that directly applying the PG algorithm \cite{birgin2000nonmonotone} to solve problem \eqref{eq.prob1} is much slower than random CD. Fig.~\ref{fig_time} shows that ideal PG is more efficient than ideal CD.  These observations are consistent with our optimization practice that it is better to coordinately update all variables together instead of individually (unless for very large-scale optimization problems where it might be computationally expensive to update all variables together).

In summary, the high computational efficiency of the proposed active set PG algorithm is mainly attributed to the following two factors. First, the active set selection strategy \eqref{selectionrule} is efficient, which is able to substantially reduce the dimension of the subproblems (compared to the original problem). Second, it is important to choose an appropriate algorithm for solving the subproblems defined over the variables in the active set and the PG algorithm \cite{birgin2000nonmonotone} turns out to be a good option (which is obviously much better than the state-of-the-art CD algorithm \cite{zhilin_icc2019,Haghighatshoar2018}).

\section{Conclusions}

Scalable and efficient joint data and activity detection is essential for massive random access in mMTC. In this paper, we propose a novel active set PG algorithm that carefully exploits the sporadic nature of the device traffic. The proposed algorithm is much more efficient than the existing state-of-the-art algorithms (in terms of the CPU time). We have observed from simulation results that several first-order algorithms can find the same (global) solution of the nonconvex joint detection problem. It will be interesting to obtain some theoretical guarantees for this observation.
%

\newpage
\bibliographystyle{IEEEtran}
\bibliography{chenbib20210204}

\begin{thebibliography}{10}
\providecommand{\url}[1]{#1}
\csname url@samestyle\endcsname
\providecommand{\newblock}{\relax}
\providecommand{\bibinfo}[2]{#2}
\providecommand{\BIBentrySTDinterwordspacing}{\spaceskip=0pt\relax}
\providecommand{\BIBentryALTinterwordstretchfactor}{4}
\providecommand{\BIBentryALTinterwordspacing}{\spaceskip=\fontdimen2\font plus
\BIBentryALTinterwordstretchfactor\fontdimen3\font minus
  \fontdimen4\font\relax}
\providecommand{\BIBforeignlanguage}[2]{{%
\expandafter\ifx\csname l@#1\endcsname\relax
\typeout{** WARNING: IEEEtran.bst: No hyphenation pattern has been}%
\typeout{** loaded for the language `#1'. Using the pattern for}%
\typeout{** the default language instead.}%
\else
\language=\csname l@#1\endcsname
\fi
#2}}
\providecommand{\BIBdecl}{\relax}
\BIBdecl

\bibitem{Bockelmann2016}
C.~Bockelmann, N.~Pratas, H.~Nikopour, K.~Au, T.~Svensson, C.~Stefanovic,
  P.~Popovski, and A.~Dekorsy, ``Massive machine-type communications in 5{G}:
  Physical and {MAC}-layer solutions,'' \emph{IEEE Commun. Mag.}, vol.~54,
  no.~9, pp. 59--65, Sept. 2016.

\bibitem{ChenX2020}
X.~{Chen}, D.~W.~K. {Ng}, W.~{Yu}, E.~G. {Larsson}, N.~{Al-Dhahir}, and
  R.~{Schober}, ``Massive access for {5G} and beyond,'' \emph{IEEE J. Sel.
  Areas Commun. (to appear)}, 2020.

\bibitem{Liu2018b}
L.~Liu, E.~G. Larsson, W.~Yu, P.~Popovski, {\v C}.~Stefanovi{\'c}, and
  E.~de~Carvalho, ``Sparse signal processing for grant-free massive
  connectivity: A future paradigm for random access protocols in the internet
  of things,'' \emph{IEEE Signal Process. Mag.}, vol.~35, no.~5, pp. 88--99,
  Sept. 2018.

\bibitem{Senel2018}
K.~Senel and E.~G. Larsson, ``Grant-free massive {MTC}-enabled massive {MIMO}:
  A compressive sensing approach,'' \emph{IEEE Trans. Commun.}, vol.~66,
  no.~12, pp. 6164--6175, Dec. 2018.

\bibitem{Liu2018}
L.~Liu and W.~Yu, ``Massive connectivity with massive {MIMO} ---{P}art {I}:
  Device activity detection and channel estimation,'' \emph{IEEE Trans. Signal
  Process.}, vol.~66, no.~11, pp. 2933--2946, June 2018.

\bibitem{Chen2018}
Z.~Chen, F.~Sohrabi, and W.~Yu, ``Sparse activity detection for massive
  connectivity,'' \emph{IEEE Trans. Signal Process.}, vol.~66, no.~7, pp.
  1890--1904, Apr. 2018.

\bibitem{liu2021efficient}
\BIBentryALTinterwordspacing
L.~Liu and Y.-F. Liu, ``An efficient algorithm for device detection and channel
  estimation in asynchronous {IoT} systems,'' in \emph{Proc. IEEE Int. Conf.
  Acoustics, Speech, Signal Process. (ICASSP), Toronto, Canada}, 2021.
  [Online]. Available: \url{https://arxiv.org/abs/2010.09979}
\BIBentrySTDinterwordspacing

\bibitem{zhilin_icc2019}
Z.~{Chen}, F.~{Sohrabi}, Y.-F. {Liu}, and W.~{Yu}, ``Covariance based joint
  activity and data detection for massive random access with massive {MIMO},''
  in \emph{Proc. IEEE Int. Conf. Commun. (ICC), Shanghai, China}, May 2019, pp.
  1--6.

\bibitem{ChenZ2020}
\BIBentryALTinterwordspacing
Z.~Chen, F.~Sohrabi, Y.-F. Liu, and W.~Yu, ``Phase transition analysis for
  covariance based massive random access with massive {MIMO},'' 2020. [Online].
  Available: \url{https://arxiv.org/abs/2003.04175}
\BIBentrySTDinterwordspacing

\bibitem{Haghighatshoar2018}
S.~Haghighatshoar, P.~Jung, and G.~Caire, ``Improved scaling law for activity
  detection in massive {MIMO} systems,'' in \emph{Proc. IEEE Int. Symp. Inf.
  Theory (ISIT), Vail, CO, USA}, June 2018, pp. 381--385.

\bibitem{Fengler2019}
\BIBentryALTinterwordspacing
A.~Fengler, G.~Caire, P.~Jung, and S.~Haghighatshoar, ``Massive {MIMO}
  unsourced random access,'' 2019. [Online]. Available:
  \url{http://arxiv.org/abs/1901.00828}
\BIBentrySTDinterwordspacing

\bibitem{Shao2020a}
\BIBentryALTinterwordspacing
X.~Shao, X.~Chen, D.~W.~K. Ng, C.~Zhong, and Z.~Zhang, ``Covariance-based
  cooperative activity detection for massive grant-free random access,'' 2020.
  [Online]. Available: \url{https://arxiv.org/abs/2008.10155}
\BIBentrySTDinterwordspacing

\bibitem{Jiang2020a}
D.~{Jiang} and Y.~{Cui}, ``{ML} estimation and {MAP} estimation for device
  activities in grant-free random access with interference,'' in \emph{Proc.
  IEEE Wireless Commun. Netw. Conf. (WCNC)}, 2020, pp. 1--6.

\bibitem{Dong2020}
\BIBentryALTinterwordspacing
J.~Dong, J.~Zhang, Y.~Shi, and J.~H. Wang, ``Faster activity and data detection
  in massive random access: A multi-armed bandit approach,'' 2020. [Online].
  Available: \url{https://arxiv.org/abs/2001.10237}
\BIBentrySTDinterwordspacing

\bibitem{Wipf2007}
D.~P. Wipf and B.~D. Rao, ``An empirical {B}ayesian strategy for solving the
  simultaneous sparse approximation problem,'' \emph{IEEE Trans. Signal
  Process.}, vol.~55, no.~7, pp. 3704--3716, July 2007.

\bibitem{yang2018sparse}
Z.~Yang, J.~Li, P.~Stoica, and L.~Xie, ``Sparse methods for
  direction-of-arrival estimation,'' in \emph{Academic Press Library in Signal
  Processing}.\hskip 1em plus 0.5em minus 0.4em\relax Elsevier, 2018, vol.~7,
  pp. 509--581.

\bibitem{birgin2000nonmonotone}
E.~G. Birgin, J.~M. Mart{\'\i}nez, and M.~Raydan, ``Nonmonotone spectral
  projected gradient methods on convex sets,'' \emph{SIAM J. Optim.}, vol.~10,
  no.~4, pp. 1196--1211, 2000.

\bibitem{bb88}
J.~Barzilai and J.~M. Borwein, ``Two-point step size gradient methods,''
  \emph{IMA J. Numer. Anal.}, vol.~8, no.~1, pp. 141--148, 1988.

\bibitem{dai2005projected}
Y.-H. Dai and R.~Fletcher, ``Projected {B}arzilai-{B}orwein methods for
  large-scale box-constrained quadratic programming,'' \emph{Numer. Math.},
  vol. 100, no.~1, pp. 21--47, 2005.

\end{thebibliography}

\end{document}